\documentclass[smallextended]{svjour3}
\smartqed  
\usepackage[fleqn]{amsmath}
\usepackage{graphicx}
\usepackage{url}
\usepackage{graphicx}
\usepackage{enumerate}
\journalname{Astronomy Letters}
\begin{document}

\title{Masses of the Main Asteroid Belt and the Kuiper Belt
from the Motions of Planets and Spacecraft}

\titlerunning {Masses of the Main Asteroid Belt and the Kuiper Belt}

\author{E. V. Pitjeva
\and N. P. Pitjev}

\authorrunning {E. V. Pitjeva, N. P. Pitjev}

\institute{E. Pitjeva \at
Institute of Applied Astronomy of Russian Academy of Sciences,
Kutuzov Quay~10, 191187 St.Petersburg, Russia \\
          \email{evp@iaaras.ru}
          \and
           N. Pitjev \at
St.Petersburg State University, Universitetski pr. 28, Petrodvoretz,
198504, Russia; Institute of Applied Astronomy of Russian Academy of Sciences,
Kutuzov Quay~10, 191187 St.Petersburg, Russia; \\
\email{ai@astro.spbu.ru}\\
}
\date{Received: 13 April 2017 / Accepted: 08 May2018
}

\maketitle

\begin{abstract}

 -- Dynamical mass estimates for the main asteroid belt and the trans-Neptunian Kuiper belt have
been found from their gravitational influence on the motion of planets. Discrete rotating models consisting of moving material points have been used to model the total attraction from small or as yet undetected bodies
of the belts. The masses of the model belts have been included in the set of parameters being refined and
determined and have been obtained by processing more than 800 thousand modern positional observations
of planets and spacecraft. We have processed the observations and determined the parameters based on
the new EPM2017 version of the IAA RAS planetary ephemerides. The large observed radial extent of the
belts (more than 1.2 au for the main belt and more than 8 au for the Kuiper belt) and the concentration
of bodies in the Kuiper belt at a distance of about 44 au found from observations have been taken into
account in the discrete models. We have also used individual mass estimates for large bodies of the belts
as well as for objects that spacecraft have approached and for bodies with satellites. Our mass estimate
for the main asteroid belt is $(4.008 \pm 0.029)\cdot 10^{-4} \ m_{\oplus}$ (3$\sigma$). The bulk of the Kuiper belt objects are in the
ring zone from 39.4 to 47.8 AU. The estimate of its total mass together with the mass of the 31 largest
trans-Neptunian Kuiper belt objects is  $(1.97 \pm 0.030)\cdot 10^{-2} \ m_{\oplus}$  (3$\sigma$), which exceeds the mass of the main
asteroid belt almost by a factor of 50. The mass of the 31 largest trans-Neptunian objects (TNOs) is only about 40\% of the total one.


\end{abstract}

\section*{\bf Introduction}
The Solar system has a complex structure that
includes large bodies and a lot of small bodies moving under the action of mutual attraction. Since the accuracy of modern radio observations is a few meters, allowance for the gravitational influence of even comparatively small objects is required to properly represent the motion of bodies in the Solar system.

The attention of Solar system researchers to the
main asteroid belt and the Kuiper belt has steadily
increased in the last decades. Most of the asteroids
that dangerously approach the Earth are associated
with the main belt, and the prediction of their appearance,
their sizes and number enters into the general
important task to prevent dangerous collisions
with the Earth. The interest in the distant Kuiper
belt is maintained with the discovery of new trans-
Neptunian objects (TNOs) and a refinement of their
distribution in distances and sizes. This belt shows a varied structure of orbits, raising questions about its formation, the composition of the primordial material, and the early history of the Solar system.

The system of differential equations of motion used
for the construction of modern numerical planetary
ephemerides contains the equations for the Sun, the
planets, the Moon, large asteroids, and TNOs. The
masses of smaller bodies are poorly known, but they
account for about 5 - 10\% of the total mass of the
main asteroid belt and $\sim 60\%$  of the mass of the Kuiper
belt, and their gravitational attraction should be taken
into account. Here we used two-dimensional discrete
models to estimate the masses of the belts. The
masses of both rings were determined by analyzing
the motion of planets (dynamical method) from
spacecraft radio data. The position of the Solar system barycenter during the calculations was controlled
and remained unchanged.

In this paper we are interested in the masses of the
main asteroid belt and the Kuiper belt in the context
of a more adequate allowance for the gravitational
influence of a large number of bodies concentrated in
the belts on the motion of major planets. A proper
allowance for the combined influence of numerous
bodies in the belts is needed to construct a more
accurate dynamical model of the Solar system and
to obtain accurate planetary ephemerides. It is possible
to properly determine the masses of the belts
from their gravitational influence on the motion of
primarily the planets closest to them. Highly accurate
observational data on the positions of spacecraft in
the vicinity of planets play a crucial role in refining the
parameters of the belts. In particular, using Saturn's
accurate observations obtained from Cassini radio
observations is important for the Kuiper belt. To refine
the masses of the belts, we used updated databases
for the main belt and TNOs, the observations of planets
and spacecraft, and the new EPM2017 version of
the IAA RAS planetary ephemerides.

\section*{EPM2017 PLANETARY EPHEMERIDES}
\noindent
The mass estimates for the Kuiper belt in this
paper are based on the processing of observations for
the latest version of the IAARAS EPM (Ephemerides
of Planets and the Moon) ephemerides, EPM2017.
The EPM2017 ephemerides contain the barycentric
coordinates and velocities of the Sun, the eight
planets, the dwarf planet Pluto, the three largest
asteroids (Ceres, Pallas, Vesta), and four TNOs (Eris,
Haumea, Makemake, Sedna) as well as the lunar
libration parameters and the TT--TDB time scale
difference. The ephemerides span a time interval
of more than 400 years (1787--2214). In the EPM
ephemerides the equations of motion for the Sun,
the Moon, and the planets obey the relativistic
Einstein--Infeld--Hoffmann equations in the inertial
barycentric frame and the TDB time scale (Pitjeva
and Pitjev 2014) with additional perturbations from
the largest asteroids and TNOs, the asteroid and
Kuiper rings, and the solar oblateness. The following
bodies treated as material points were included in the
dynamical system: the planets, the Moon, Pluto, the
301 largest asteroids, and the 30 largest TNOs. The
Newtonian gravitational accelerations are assumed
to be negligible in the case where both interacting
bodies are neither the Sun, nor a planet, nor one of
the following bodies: the Moon, Pluto, Ceres, Pallas,
Vesta, Iris, and Bamberg. In other words, for 16
main objects the equations of motion include all of
the mutual perturbations, while for the remaining
asteroids and TNOs only the Newtonian mutual 
perturbations between them and the planets, the
Moon, and the Sun are calculated.

The major modifications compared to the previous
EPM versions concern the following:
\begin{itemize}
\itemsep 1mm
\item
The new revised version of the software package
ERA-8 (Pavlov and Skripnichenko 2015)
using the C and Racket programming languages,
the SQLite database, and the SOFA
(Standards Of Fundamental Astronomy) library
of astronomical calculations is employed.
\item A new model of the Moon is orbital and rotational
motion (Pavlov et al. 2016) based on the
equations of lunar motion of the JPL DE430
ephemerides (Folkner et al. 2014) and recommended
by geophysical and geodynamic models
was constructed. The change in the gravitational
potential of the Moon as a result of
tidal and rotational deformations, the torque of
the Moon's liquid core, and the dissipation of
energy under core friction against the crust are
taken into account in the realization.

\item The model for the motion of planets and other
objects in the Solar system in which the models
of solar motion and the barycenter equations
were refined was updated. Two-dimensional
discrete rotating rings of the main asteroid belt
and the Kuiper belt were added to the dynamical
model of the Solar system and 30 TNOs
were included in the joint integration.

\item The additional relativistic Lense -- Thirring accelerations
dependent on the solar rotation
were included in the general model of motions
of Solar system bodies (http://iaaras.ru/en/dept/ephemeris/epm/2017/).

\item The number of observations increased -- we
used a total of $\sim $­ 800 thousand positional
observations of planets and spacecraft (1913--2015) and lunar laser observations (LLOs)
(1970--2016), including the new infrared LLOs
and the Cassini and Messenger data.

\item The data on a number of asteroids and TNOs
were updated and refined, in particular, we used
the new masses deduced in the Dawn spacecraft
encounters with Vesta and Ceres as well
as for binary asteroids or those and TNOs with
satellites.
\end{itemize}

\subsection*{\it Refining the Model of Solar Motion
and the Equations of the Solar System Barycenter}
The equations of motion in the dynamical EPM
model are specified in the Barycentric Celestial Reference
System (BCRS) whose origin is at the Solar system barycenter. The relativistic barycenter of the
point masses $\mu_i$ of bodies is defined as

$$  \bf b = {{\sum_i {\mu_i}^*{\bf r}_{i}}\over{\sum_i {\mu_i}^*}}, $$
where  ${\bf r}_{i}$ is the position of body i and ${\mu_i}^*$  is its relativistic
gravitational parameter:
 
 $${\mu_i}^* = \mu_i \left(1 + {1\over{2c^2}}{\dot r_i}^2 - {1\over{2c^2}}
\sum_{j \ne i}{{\mu_j}\over{r_{ij}}} \right)$$
 
The notation:

	$\mu_i= Gm_i$, where  G the gravitational constant,  $m_i$ is the mass of body i; for the planets, except the
Earth, and other objects the masses $m_i$ also include
the masses of their satellites;

c -- the speed of light;

	$ r_{ij}=|r_{ij}|=|r_j-r_i| $.
	
	The initial coordinates and velocities of all bodies
in the dynamical system are shifted so that the conditions
$\bf b = 0$ and $\bf \dot b = 0$ are fulfilled. This is achieved
by solving the following system of equations by an
iterative method: 

   $$ \sum_i {\mu_i}^*{\bf r}_{i} = {\bf 0}   ,      $$
   $$ \sum_i \left({\dot \mu_i}^*{\bf r}_{i} + {\mu_i}^*{\bf \dot r}_i\right) = {\bf 0} , $$

where 	${\bf \dot r}_i$ is the velocity vector of body i. 

An improvement realized in EPM2016 and subsequent
ephemerides is the inclusion of the term ${\dot \mu_i}$ that
was neglected in EPM2015 and earlier ephemerides.

The Sun is involved in calculating the accelerations,
along with other bodies, with the mutual gravitational
influence of the Sun, the planets, large asteroids,
TNOs, and the introduced rings for modeling
the total attraction from small bodies of the main
asteroid belt and the Kuiper belt being taken into
account.

A peculiarity of the BCRS is the uncertainty of
its axes. The initial positions of the axes are close
to the J2000 coordinate system. When the observations
are processed, the rotation correction parameters
 $\epsilon_x, \epsilon_y$ and $\epsilon_z$ are refined 
for the orientation
of the final ephemerides in the International Celestial
Reference Frame ICRF2 using special spacecraft
observations on the background of quasars whose
coordinates are given in ICRF2. The accuracy of
the BCRS orientation in the ICRS is about 0.2 mas
(Pitjeva et al. 2017).

The roundoff errors in the numerical integration
lead to a slight drift of the Solar system barycenter in
the model from the initial position at the coordinate
origin. However, this drift is insignificant (in the
interval from 1900 to 2020 it is less than 0.02 mm)
and virtually negligible in problems of ephemeris astronomy.
\subsection*{\it Modeling the Gravitational Influence of the Main Asteroid Belt and the Kuiper Belt}
When modern highly accurate planetary ephemerides
are constructed, the system of equations that
includes the equations of motion for the Sun, the
Moon, the planets, and the 301 largest asteroids
(with EPM1999) and largest TNOs (21 with with
EPM2008 and 30 with EPM2015) is integrated.
Such a direct inclusion in the general system of integration
requires a thorough and careful refinement
of the initial data and the masses of the asteroids and TNOs.

Dynamical mass estimates are available for a
number of asteroids that spacecraft have approached,
for asteroids and TNOs with satellites, and for large
asteroids from their perturbations on Mars and the
Earth determined by radar observations. Individual
mass estimates for asteroids and TNOs can also be
obtained from their density estimates by taking into
account their diameters and taxonomic class, but
with a considerably larger error up to $5 \cdot 10^{-12} \ M_{Sun}$.
The total number of asteroids with estimated masses
included in the EPM integration is 301 (Pitjeva
and Pitjev 2016). For 46 asteroids the masses
were inferred from individual perturbations. For
30 asteroids the masses were inferred from their
perturbations of planets by processing the radar
data from Martian spacecraft and landers. The
masses of Eros, Vesta, and Ceres were estimated very
accurately from the data of the NEAR and DOWN
spacecraft that investigated these asteroids. For 13
asteroids with satellites different authors obtained
quite accurate estimates of their masses. For the
remaining 255 asteroids the masses were determined
from their diameters and density estimates for three
taxonomic classes obtained when improving the EPM ephemerides. Thus, we found the total mass of 301 largest asteroids to be 

$M_{301aster} = (11.531 \pm 0.071) \times 10^{-10} M_{Sun} \qquad ( 3 \sigma)$.

\vskip 0.5cm
The Pluto--Charon system was included in the
model to construct the ephemerides from the first
EPM versions. At the same time, this system also
belongs to the Kuiper belt. Mass estimates are available
for a number of Kuiper belt objects around which
satellites were detected. For such 11 large objects
(Table 1) the masses taken from various astronomical
data sites were determined with a reasonably good
accuracy.
\begin{table}[h]

\centering
{{\bf Table 1.} Mass estimates for large Kuiper belt objects with satellites}

\vspace{2mm}\begin{tabular}{l|c|l|c} \hline\hline
 TNO & Mass, $m_{\oplus}$ & TNO & Mass, $m_{\oplus}$ \\
\hline\
 Eris & $(28.0 \pm 0.33)\times 10^{-4}$ & Orcus & $(1.059 \pm 0.008)\times 10^{-4}$ \\ 
 Pluto+Charon &  $(24.473 \pm 0.113)\times
 10^{-4}$ & 2003 $AZ_{84}$ & $(0.69 \pm 0.33)\times 10^{-4}$ \\ 
 Haumea & $(6.71 \pm 0.067\times 10^{-4}$ & Salacia & $(0.73 \pm 0.03)\times 10^{-4}$ \\
 Makemake & $(4.35 \pm 0.84)\times 10^{-4}$ & Varda  & $(0.444 \pm 0.011)\times 10^{-4}$ \\
 2007 $OR_{10}$ & $(6.3 \pm 4.5)\times 10^{-4}$ & 2002 $UX_{25}$ & $0.209 \pm 0.005)\times 10^{-4}$ \\
 Quaoar &  $(1.67 \pm 0.17)\times 10^{-4}$ &  \\ 
\hline
\end{tabular}
\end{table}

For the remaining 20 of the 31 largest TNOs the
masses were estimated much more poorly; they were
calculated from their diameters and densities. The
total mass of the 31 individually estimated largest
TNOs, including the Pluto--Charon system, is
\begin{equation}\label{f-7}
 M_{31TNO} = 0.0086 \pm 0.0017  \ m_{\oplus}  \qquad  (3 \sigma).     
\end{equation}
The masses of smaller bodies are poorly known,
but they are variously estimated to be about 5-10\%
of the total mass of the main asteroid belt and 60\% of
the mass of the Kuiper belt. Their combined gravitational
pull on the planets was also taken into account
using a special model.

Allowance for the total additional attraction from
the asteroids neglected individually in the joint numerical
integration was first proposed by G.A. Krasinsky.
The attraction from these objects was modeled
by the attraction from a material circular ring located in the plane of the ecliptic with a uniform distribution of matter. The formulas for the disturbing force of the asteroid ring are given in Krasinsky et al. (2002). The mass of the ring $M_r$ and its radius $R_r$ were included
in the set of parameters being improved. This model
of a homogeneous ring was used in constructing the
EPM2004 ephemerides (Pitjeva 2005). However, on
the one hand, the asteroid belt is quite wide (more
than 1 AU) and, on the other hand, there is a strong
correlation between two parameters of the ring, its
mass $M_r$ and radius $R_r$, that does not allow them to be estimated with a good accuracy. Therefore, when
constructing the EPM2013 ephemerides (Pitjeva and
Pitjev 2014), we decided to pass to modeling the small bodies of the main asteroid belt by a two-dimensional
ring with its radii specifying the boundaries of the
model ring within which the bulk of the belt bodies
are located and which correspond to the two main
1 : 2 and 1 : 4 orbital resonances with the motion of Jupiter.

The situation with the Kuiper belt is similar. First
we modeled the attraction from small TNOs by a one-dimensional ring located in the plane of the ecliptic
with a radius of 43 AU and its estimated mass
(Pitjeva 2010a, 2010b). In 2017, when constructing
EPM2016 (Pitjeva et al. 2017), we used a two-dimensional rings of TNOs with the radii attributable
to the 2 : 3 and 1 : 2 orbital resonances with the motion of Neptune.

Finding the gravitational potential and its derivatives for a flat material two-dimensional homogeneous ring leads to expressions that include complete elliptic integrals of the first, second, 
and third kinds. The calculation can be reduced to calculating the
values of the hypergeometric function of the corresponding parameters by first applying the Landen transformation. The expressions for the potential and
accelerations of the two-dimensional ring via the hypergeometric functions of four arguments were 
given in Pitjeva and Pitjev (2014, 2016). The formulas for the accelerations differ for points on the inside (closer to the Sun) and the outside (farther from the Sun) with respect to the two-dimensional circular ring. In 2017 they were first used for the Kuiper belt
(Pitjeva et al. 2017).

\subsection*{\it A Discrete Rotating Model for the Belts}
The one- and two-dimensional models used in
the previous EPM2004--EPM2016 versions of the
ephemerides to find the total gravitational attraction
from small bodies in the belts had a disadvantage.
When the mutual attraction between the planet and
the belt was taken into account, the introduced ring
model should have shifted as a whole, which does
not correspond to the actual interaction between the
planet and moving bodies in the belts. To avoid this
disadvantage, in this paper we used a discrete rotating model. Note that, as in previous papers, we modeled not the belt and the motion in it, but its gravitational
influence on the planets. The discrete models are a
system of material points with equal masses located
in the plane of the ecliptic and moving in the same
(prograde) direction as the planets. The material
points in the model do not interact between themselves. The gravitational attraction occurs between each of
them with the Sun and planets. In the model
the material points were arranged at the initial time uniformly on three circular lines with initial velocities corresponding to circular motion.

For the main asteroid belt the extreme lines of the
model correspond to the 2.06 and 3.27 au boundaries
within which the bulk of the asteroids are located.
Outside these boundaries the number of asteroids
drops sharply. These distances correspond to the 1 : 2 and 1 : 4 orbital resonances with the mean motion of Jupiter. The bulk of the small asteroids, asteroid fragments, and dust of the main belt are located in the same zone, but the fraction of all of them in the mass of the main belt is small, $\sim$ 5--10\%. Sixty material points were located on each of the extreme lines and the middle line with a radius of 2.67 au in the model.

For the Kuiper belt the extreme lines with the system of material points corresponded to the boundaries of the densest, ``main'' part (De Sanctis et al. 2001) of the belt ($R_1$ = 39.4 au and $R_2$ = 48.7 au). These distances correspond to the 2 : 3 and 1 : 2 orbital resonances with the mean motion of Neptune. Forty points were located on each of these lines. The third line $R_m$= 44 au corresponded to the observed crowding of objects near 44 au, the belt ``kernel''
(Petit et al. 2001; Bannister et al. 2016), and the
number of model points on it was 80, twice as large
as that on the extreme lines.

The total masses of each model were the parameters
that were determined by processing the observational
data.

\subsection*{\it Observational Data}

The number of highly accurate observations on
which the next EPM versions are based increases
steadily, and the total number of observations used
in the current EPM2017 version of the planetary
ephemerides is more than 800 thousands. In this
case, not the individual spacecraft measurements
themselves, but the normal points into which the
observations on one spacecraft revolution were combined are used, because these observations are correlated between themselves. However, the observations of Martian landers were not combined into normal points, because they were also used to study the rotation of Mars in addition to the determination of other parameters. 
\begin{table}[h]

\centering
{{\bf Table 2.}  The observations used to improve EPM2017 and to estimate the parametersÍàáëþäåíèÿ, èñïîëüçîâàííûå äëÿ óëó÷øåíèÿ EPM2017 è îöåíèâàíèÿ ïàðàìåòðîâ}

\vspace{2mm}\begin{tabular}{c|c|c|c|c}
\hline \hline
   & \multicolumn{2}{c|}{Radio observations} & \multicolumn{2}{c}{Optical observations} \\
   \hline
Planet & interval of
 & number of & interval of  & number  of  \\
  & observations & norm. points & observations & observations \\
\hline
Mercury & 1964--2015 & 1556  & --- & --- \\
Venus   & 1961--2013 & 3621  & --- & --- \\
Mars    & 1965--2014 & 46441 & --- & --- \\
Jupiter + 4 sat. & 1973--1997 & 51 & 1914--2013 & 14866 \\
Saturn + 7 sat.  & 1979--2014 & 171 & 1913--2013 & 16455 \\
Uranus + 4 sat.  & 1986       & 3 & 1914--2013 & 12550 \\
Neptune + 1 sat. & 1989       & 3 & 1913--2013 & 12404 \\
Pluto          & ---        & --- & 1914--2013 & 16674 \\
\hline
Total & 1961--2015 & 51846 & 1913--2013 & 72049 \\
\hline
\end{tabular}
\end{table}

\vspace{-3mm}
\noindent
Table 2 gives the number of normal points for radar data and the number of observations for optical data. Only highly accurate radio measurements that span a time interval of more than half a century are currently used in constructing the ephemerides of the inner planets. The optical observations are several orders of magnitude less accurate and were not used for these planets.

The techniques for determining the parameters of
planetary ephemerides by processing measurements
of various types, from classical meridian observations of planets and their satellites to modern radio observations
of planets and spacecraft, are described in
Pitjeva (2005, 2013). In this paper we used the optical
observations from 1913 up until all the accessible
modern 2015 observations. The accuracies of modern
optical CCD observations reach a few hundredths of
an arcsecond; spacecraft trajectory observations are
much more accurate: 1--2 m for the inner planets
and about 20 m for the Cassini spacecraft (Hees
et al. 2014). The optical observations with an
acceptable accuracy span an interval slightly longer
than one revolution for Uranus and slightly more
than half of the complete revolution around the Sun
for Neptune. From the radio measurements for
Uranus and Neptune there is only one 3D point for
each of the planets obtained during the Voyager-2 flyby. Most of the observations were taken from
the database of the NASA Jet Propulsion Laboratory
(JPL) created by M. Standish and being
continued and maintained at present by W. Folkner:
https://ssd.jpl.nasa.gov/?eph$\_$data.
 These data were
supplemented by the Russian radar observations
of planets (1961--1995): http://iaaras.ru/en/dept/ephemeris/observations/, the Pulkovo Observatory
optical data and the revised Lowell Observatory data
(Buie and Folkner 2015), the new CCD data of the
Brazilian Pico dos Diasa Observatory (Benedetti-Rossi et al. 2014), and the Venus Express (VEX) and
Mars Express (MEX) data retrieved by courtesy of
A. Fienga, 

http://www.geoazur.fr/astrogeo/?href= observations/base. 

After the processing of all observations and the
refinement of the EPM2017 parameters, Figs. 1--7
present the residuals of the ranging from the Earth
to the planet for the observations of the Messenger,
VEX, Mars Global Surveyor (MGS), Odyssey, MEX,
Mars Reconnaissance Orbiter (MRO), and Cassini
spacecraft revolving about Mercury, Venus, Mars,
and Saturn.

It should be noted that for the observations near
conjunctions of the planet with the Sun the scatter of residual increases significantly due to the influence of
the solar corona (despite its reduction in the observations);
therefore, these observations were removed.
The root-mean-square (rms) deviations of the residuals ($\sigma$)
of the method, wrms, for the corresponding
sets of observations are given in the captions to the
figures. The rms deviations are 0.7 m for the new
Messenger observations, about 1 m for the Martian
spacecraft, and 21.2 m for the Cassini spacecraft near
Saturn. For the VEX spacecraft near Venus (2010--
2012) the spacecraft orbit after 2010 passed near the
upper layers of its atmosphere, occasionally touching
it. Therefore, the scatter of residuals for the VEX observations
is slightly larger than that for the Martian
spacecraft, being about 3 m. Despite the fact that the
agreement of the Martian observations became more
difficult, because the interval of observations of the
Martian spacecraft increased compared to EPM2011
by 3--4 years, the rms deviations for the Martian
MGS, Odyssey, and MRO spacecraft nevertheless
decreased by 10--20\% due to the refinement of the
models of planetary motions. The graphs of the residuals
of spacecraft observations and the amplitudes of
their scatter are close to those for the JPL NASA
DE 430 ephemerides (Folkner et al. 2014), while the
characteristics in Fig. 1 for the Messenger spacecraft
are similar to the graph from Park et al. (2017).

\vspace{-7mm}\begin{figure}[h]
\includegraphics[scale=1.8]{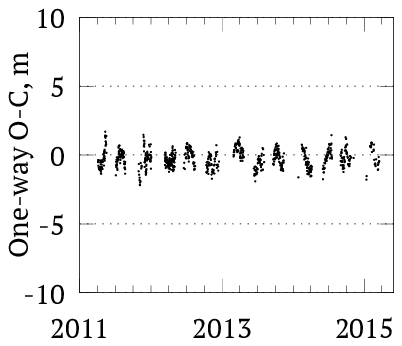}
\vspace{-3mm}\caption{O--C of the Messenger ranging
(in one direction) calculated from the EPM2017 ephemerides, $\sigma$ = 0.7 m.}

\vspace{-3mm}
\includegraphics[scale=1.8]{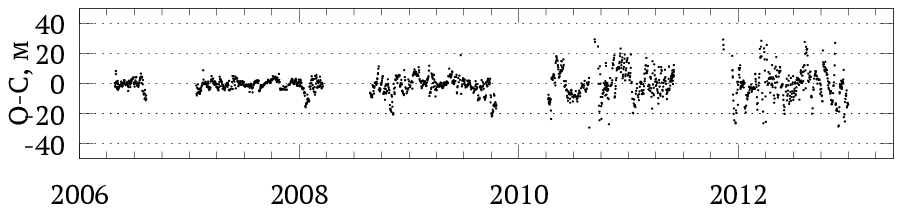}
\vspace{-7mm}\caption{O--C of VEX ranging
(in one direction) calculated from the EPM2017 ephemerides, $\sigma$ = 2.98 m.}

\vspace{-3mm}\includegraphics[scale=1.8]{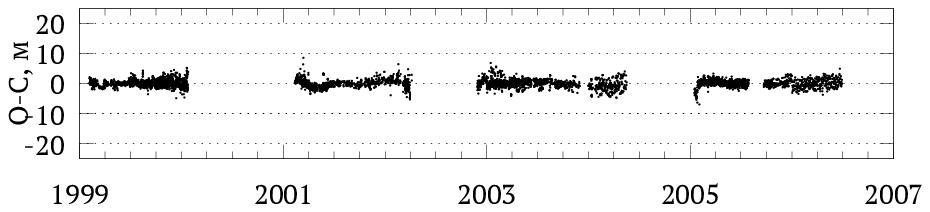}
\vspace{-7mm}\caption{O--C of MGS ranging
(in one direction) calculated from the EPM2017 ephemerides, $\sigma$ = 1.17 m.}

\vspace{-3mm}\includegraphics[scale=1.8]{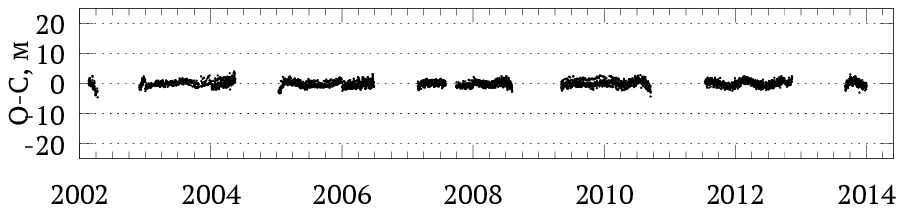}
\vspace{-7mm}\caption{O--C of Odyssey ranging
(in one direction) calculated from the EPM2017 ephemerides, $ \sigma $= 0.95 m.}
\end{figure}

\begin{figure}[ht]
\vspace{-3mm}\includegraphics[scale=1.8]{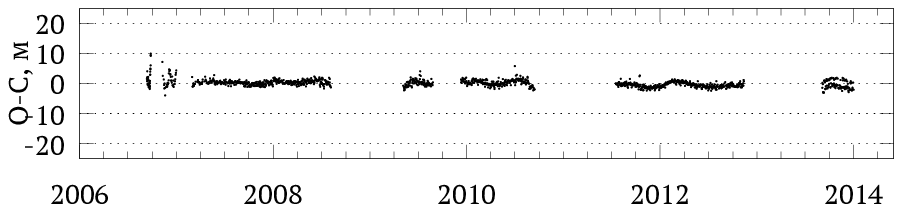}
\vspace{-7mm}\caption{O--C of the MRO ranging (in one direction) calculated from the EPM2017 ephemerides, $\sigma$= 0.95 m.}


\vspace{-3mm}\includegraphics[scale=1.8]{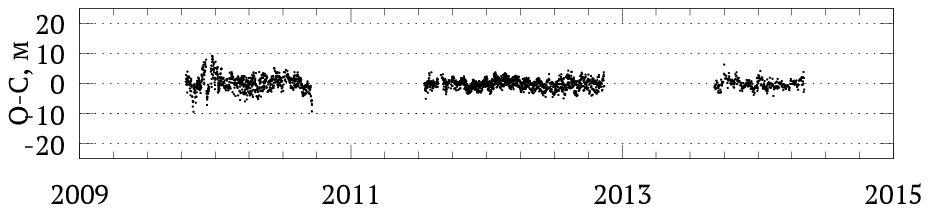}
\vspace{-7mm}\caption{O--C of the MEX ranging (in one direction) calculated from the EPM2017 ephemerides, $\sigma$ = 1.5 m.}


\vspace{-3mm}\includegraphics[scale=1.8]{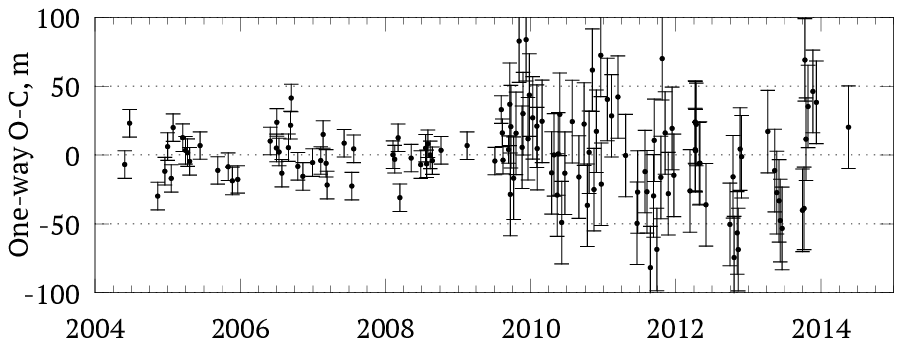}
\vspace{-7mm}\caption{O--C of the Cassini ranging (in one direction) calculated from the EPM2017 ephemerides, $\sigma$ = 20.2 m.}
\end{figure}
\noindent

\clearpage

\subsection*{\it Accuracies of the Orbital Elements of the EPM Ephemerides}
Concurrently
 with the refinement of all parameters
of the EPM2017 ephemerides, we also obtained their
formal errors. Table 3 gives the formal standard errors of the orbital elements for the planets, where $a$ is the semimajor axis, $ i $ is the orbital inclination, $\Omega$ is the
longitude of the ascending node, e is the eccenticity, $\pi$ is the longitude of perihelion, and $ \lambda $ is the mean longitude. The formal standard errors of the least squares
method (LSM) for the semimajor axes of
the inner planets are a few hundredths of a meter
(Table 3), and although, as experience shows, the
actual errors can be larger than the formal LSM errors
by an order of magnitude, the accuracy of determining
these and other orbital elements is high even if this is taken into account.

\begin{table}[th]

\centering
{{\bf Table 3.} Formal standard errors of the orbital elements for the planets calculated from the EPM2011 and EPM2017 ephemerides}

\vspace{2mm}\begin{tabular}{llcccccc}
\hline\hline
Ephemerides & Planet  & $a$  & $\sin i\cos\Omega$ & $\sin i\sin\Omega$
                          & $e\cos\pi$ & $e\sin\pi$ & $\lambda$\phantom{00} \\
  &  & [m]  & [mas] & [mas] & [mas] & [mas] & [mas] \\ \hline
EPM2011 & Mercury &  0.170  & 0.8275 & 0.5639 & 0.0907 & 0.06885 & 0.1617 \\
EPM2017 &  Mercury &  0.0015 &  0.00152 & 0.00144 & 0.00084 & 0.00024 & 0.00337 \\
\hline
EPM2011 & Venus   &  0.089 & 0.00364 & 0.00288 & 0.00033 & 0.00020 & 0.00325 \\
EPM2017 & Venus   &  0.0065 &  0.00358 & 0.00349 & 0.00014 & 0.00016 & 0.00268 \\
\noalign{\smallskip}\hline\noalign{\smallskip}
EPM2011 & Earth    &  0.131 & \phantom{0.}---   & \phantom{0.}---   & 0.00043 & 0.00017 & \phantom{0.}---    \\
EPM2017 &Earth   &  0.013 & \phantom{0.}---   & \phantom{0.}---   & 0.00008 & 0.0006 & \phantom{0.} \\
\noalign{\smallskip}\hline\noalign{\smallskip}
EPM2011 & Mars    &  0.616 & 0.00143 & 0.00115 & 0.00142 & 0.00071 & 0.00278 \\
EPM2017 & Mars    &  0.0487 &  0.00084 & 0.00093 & 0.00008 & 0.00018 & 0.00037 \\
\noalign{\smallskip}\hline\noalign{\smallskip}
EPM2011 &  Jupiter &  351 & 2.008 & 1.811 & 0.129 & 0.110 & 0.884 \\
EPM2017 & Jupiter &  372   & 1.749 & 1.629 & 0.163 & 0.133 & 1.070 \\
\noalign{\smallskip}\hline\noalign{\smallskip}
EPM2011 & Saturn  & 70.519 & 0.10792 & 0.12023 & 0.01093 & 0.00327 & 0.03434 \\
EPM2017 & Saturn  & 16.936 & 0.08176 & 0.05845 & 0.00368 & 0.00237 & 0.01732 \\
\noalign{\smallskip}\hline\noalign{\smallskip}
EPM2011 & Uranus & 30075 & 3.458 & 4.013 & 2.853 & 2.006 & 3.598 \\
EPM2017 & Uranus & 31314 & 3.574 & 3.806 & 2.716 & 2.238 & 2.833 \\
\noalign{\smallskip}\hline\noalign{\smallskip}
EPM2011 & Neptune & 270853 & 2.673 & 5.202 & 5.554 & 13.558 & 12.363 \\
EPM2017 & Neptune & 288035 & 3.769 & 5.381 & 5.791 & 14.386 & 12.536 \\
\noalign{\smallskip}\hline\noalign{\smallskip}
EPM2011 & Pluto &  2022765 & 2.759 & 10.021 & 43.896 & 31.381 & 18.215 \\
EPM2017 & Pluto &  790006 & 0.758 &  3.657 & 17.671 & 12.473 & 6.022 \\
\hline
\end{tabular}           	
\end{table}

In addition, Table 3 shows, for comparison, the
formal standard errors of the orbital elements for the
planets calculated from the EPM2011 ephemerides.
It should be noted that these standard errors were calculated
for the EPM2011 and EPM2017 ephemerides
using two different software packages, ERA-7 and
ERA-8. Therefore, these errors slightly differ even
when using the same observations. This can be
seen using Neptune as an example, for which no new
observations have been available since 2011. For
Mercury the orbital elements became much more
accurate, while their formal errors decreased by one
or two orders of magnitude after the appearance of
Messenger observations. The formal errors of the
orbital elements for Mars and the Earth decreased
noticeably (occasionally by an order of magnitude)
due to the large amount of new Odyssey, MRO, and
MEX data. For Pluto the decrease in the errors of the orbital elements by several times is explained by the
appearance of new highly accurate CCD observations
obtained at the Pico dos Dias Observatory from 1995
to 2013 and revised Lowell Observatory data (1913–
1951). For Venus the VEX mission ended in 2012 and
no new observations have appeared, just as for Jupiter and Uranus; therefore, their errors barely changed. As yet no new observations from the Juno spacecraft near Jupiter are accessible.

The progress of EPM2017 in increasing the accuracy
of planetary ephemerides is explained primarily
by the use of new highly accurate trajectory data for various spacecraft as well as by an improvement in
modeling the motions of planets and an improvement
of the dynamical model.

		\section*{\bf MASS OF THE MAIN ASTEROID BELT}
A large set of small bodies moving in nearly circular
orbits between Mars and Jupiter, 1.8 au $< r <$
3.5 au, belong to the main asteroid belt. The densest part of the belt is located in the band between the 1 : 4
(2.06 au) and 1 : 2 (3.27 au) orbital resonances
with Jupiter. It contains more than 90\% of all the
numbered asteroids. These distances may be deemed
to be, respectively, the inner and outer boundaries
of the bulk of the main belt, because the number of
asteroids drops sharply outside them.

\subsection*{\it Statistical and Dynamical Mass Estimates for the Asteroid Belt}
The first estimates of the total mass for the belt
were made by statistical methods. The distributions
in apparent magnitudes obtained from observations,
albedo estimates, deduced and empirical size distribution
functions, and estimates of the mean densities
of objects were used. For the asteroids of the main
belt the mean densities depended on the taxonomic
class to which a given asteroid was attributed. Since
the statistical estimates depend on several assumptions
and ill-defined parameters, the authors provide
no errors of the statistical mass estimates (Table 4).

The mass of the belt was grossly overestimated
in 1990 (McBride and Hughes 1990). Subsequently,
the values of the mass were reduced. The latest paper on a statistical mass estimate was published in 2012 (Vinogradova 2012).
\begin{table}[h]

\vspace{6mm}
\centering
{{\bf Table 4.} Statistical estimates of the total mass for the main asteroid belt }
 
\vspace{5mm}\begin{tabular}{l|c|c}
\hline \hline
Year & AuthorsÀâòîðû & Mass (in $ M_{\odot}$) \\
\hline 
1990 &  McBride and Hughes & ${\sim 55\cdot10^{-10}}$ \\
2001 &  Petit et al. & $15 \cdot10^{-10}$ \\
2012 & Vinogradova & $13.5 \cdot10^{-10}$ \\
\hline
\end{tabular}
\end{table}
Dynamical mass estimates for the main belt are
obtained from its gravitational influence on the motion of other bodies in the Solar system, primarily on Mars nearest to the belt. Mass estimates for 301 large
asteroids are presented in the Section ``Modeling the Gravitational Influence of the Main Asteroid Belt and the Kuiper Belt''.

The total mass of the remaining small asteroids,
asteroid fragments, and dust in the main belt was
found in the EPM ephemerides from observations
using initially a one-dimensional ring with its estimated
radius and mass (EPM2000--EPM2011) and
subsequently using a two-dimensional ring with its
radii determined by the 2.06 and 3.27 au resonances
with the motion of Jupiter and the estimated mass
(beginning with EPM2013).

The total mass of the main asteroid belt was found
as the sum of the masses of the 301 largest asteroids
and the estimated mass of the modeled asteroid ring.
Table 5 gives previous mass estimates for the main
asteroid belt. For comparison, the mass estimate for
the asteroid belt obtained for the DE 430 ephemerides
(Kuchynka and Folkner 2013) was also included in
the table. In its paper the mass of the asteroid
belt was found from the masses of 3714 asteroids
estimated in two ways. For the largest 343 asteroids
the masses were determined using Tikhonov's regularization
from the ranging measurements of Martian
spacecraft and landers; for the remaining 3371 asteroids
the mass estimates were obtained from their
diameters deduced from infrared observations and the
presumed mean density $\rho$ = 2.2 g cm$^{-3}$.
 The derived
total mass of the belt in their paper is $13.3\cdot10^{-10} M_{\odot}$, with 90\% of this mass being accounted for by 343 large asteroids.
\begin{table}[t]

\vspace{6mm}
\centering
{{\bf Table 5.} Dynamical estimates of the total mass for the main asteroid belt} 

\vspace{0mm}\begin{tabular}{l|c|c|c}
\noalign{\smallskip}
\hline
\noalign{\smallskip}
Year & Authors & Ephemerides & Mass (in $ M_{\odot}$) \\
\hline\hline 
\noalign{\smallskip}
2002 & Krasinsky et al. & EPM2000 & $(18 \pm 2)\cdot10^{-10}$ \\
2005 & Pitjeva & EPM2004 & $(15 \pm 1)\cdot10^{-10}$\\
2013 & Kuchynka and Folkner & DE 430 & $(13.3 \pm 0.2)\cdot10^{-10}$ \\
2013 & Pitjeva & EPM2011 & $(12.3 \pm 1.2)\cdot10^{-10}$\\
2014 & Pitjeva and Pitjev & EPM2013 & $(12.2 \pm 0.2)\cdot10^{-10} $\\
2017 & Pitjeva et al. & EPM2016 & $(12.245 \pm 0.187)\cdot10^{-10} $\\
2018 & Pitjeva and Pitjev & EPM2017 & $(12.038 \pm 0.0874)\cdot10^{-10} $\\
\hline
\end{tabular}
\end{table}

In our paper to estimate the mass of the remaining,
smaller asteroids, we used a discrete rotating
model with radii R$_1$ = 2.08 au and R$_2$ = 3.27 au
for the extreme lines and the middle line with radius R$_m$ = 2.66 au and 60 material points on each of the lines and a total number of moving material points of 180 (see Section ``A Discrete Rotating Model for the Belts'').

The total mass of the small asteroids, small fragments,
and dust of the main belt was found from the
discrete model to be

\noindent $ M_{discr.ring} = (0.507 \pm 0.051) \cdot 10^{-10} \ M_{sun}=  
       (0.169 \pm 0.017) \cdot 10^{-4} \ m_{\oplus} \ (3\sigma). $
       
The total mass of the main belt (in the sum with
the mass of 301 large asteroids) is

$ M_{Main} = (12.038 \pm 0.087) \cdot 10^{-10} \ M_{sun}= 
       (4.008 \pm 0.029) \cdot 10^{-4} \ m_{\oplus} \ (3 \sigma). $
       
The same mass is shown in the last row of Table 5.
All of the mass estimates in this paper are given with
an uncertainty equal to the $ 3\sigma $ standard error of the
LSM. The mass of the small bodies and dust is about
4.5\% of the total mass of the main belt. It can be
seen from Table 5 that the error in estimating the total
mass of the main asteroid belt decreases in each new
EPM version due to an increase in the amount of
observational data and an improvement of the models
used for the mass estimates of the remaining small
asteroids. The mass estimation accuracy improved
approximately by a factor of 6 as we passed from the
one-dimensional ring to the two-dimensional one and
became better for the discrete rotating ring by a factor
of 2. The representation of observations also improved.
The residuals in processing the observations
of the spacecraft near Venus and Mars are given in
Figs. 1--6.  

\section*{\bf MASS OF THE KUIPER BELT}
TheKuiper belt objects (KBOs) are divided (Jewitt
et al. 1998; Gladman 2002), given the characteristics
of their orbits, into three main dynamical classes:
classical, resonant, and scattered disk objects.

{\bf Classical KBOs} have nearly circular orbits and
relatively small eccentricities. Their nearly circular
orbits lies in the region 40--50 au from the Sun.
These objects undergo no strong influence of the major
planets; their orbits remain essentially unchanged.
They are most numerous and constitute the bulk of
the population. The plane of the Kuiper belt determined from classical objects with inclinations $|i| < 5{^\circ}$
agrees with the invariant plane of the Solar system
(Elliot et al. 2005). The main classical Kuiper belt is located between the 3 : 2 and 2 : 1 orbital resonances
with Neptune in the ring belt 39.4 au$< a <$ 47.8 au
(De Sanctis et al. 2001). There is a sharp outer edge in the distribution, the so-called Kuiper Cliff, a sharp
drop in the number of classical objects after 50 au.
To be more precise, beyond 48 au the number of
objects with sizes larger than 40 km drops sharply
(Jewitt et al. 1998; Trujillo and Brown 2001; Gladman et al. 2001; Allen et al. 2002). The outer boundary for the belt of classical objects is well-defined, and it may correspond to the edge of the primordial protoplanetary cloud. 

    There is a crowding in the distribution (``kernel'') with a concentration of orbits with semimajor axes $a \sim 44$ au, eccentricities $e \sim 0.05$, and inclinations $|i| < 5^{\circ} $
(Petit et al. 2001; Bannister et al. 2016).

Primarily the objects in the 3 : 2 (plutino, $a \sim  39.4$ au) and 2 : 1 ($a \sim  47.8$ au) orbital resonances
with the mean motion of Neptune are attributed to
{\bf resonant KBOs}, although there are some number
of bodies with different resonance ratios of the mean motions.   

 {\bf Scattered disk objects} ("wanderers"), objects
with large eccentricities and large inclinations, with
orbits extending far beyond 50 au (up to $a \sim $90~au and  $e \sim 0.5-0.6$ or larger) constitute the sparsest part. 

The formation of such a belt structure is explained
by perturbations from the planets and primarily from Neptune (Levison and Morbidelli 2003).
\subsection*{\it Statistical Mass Estimates for the Kuiper Belt}
The first mass estimates for the Kuiper belt were
obtained by statistical methods, where the parameters
of the size distribution of bodies are used, which,
in turn, are found from the distribution in apparent
magnitudes. Because of the large distance to Kuiper
belt objects, small objects are invisible even through
large telescopes, and the size distribution function
has to be extrapolated into the range of small diameters and the uncertainty of the result increases. Estimates
of the mean density are also used. As a rule, the
mean density is taken to be 1.5--2.0 g cm$^{-3}$, because the Kuiper belt objects are icy bodies that incorporate
frozen methane, ammonia, water, and carbon dioxide
surrounding the rocky interiors. Generally, the statistical
estimates have a large scatter and are based
on various, not quite reliable assumptions about the
albedo and density of belt objects. Because of their
large uncertainties, the authors publish the mass estimates
without providing any errors. Table 6 gives
the values obtained by different authors. The scatter of estimates is large; the estimates lie within the range
from 0.01 $m_{\oplus}$ to 0.2 $m_{\oplus}$.

In the suggested models of the formation of the
Solar system a significant initial mass of the belt,
$\sim 10-30m_{\oplus}$, is required for the formation of the
Kuiper belt (Stern and Colwell 1997; Delsanti and
Jewitt 2006; Kenyon 2002). The statistical mass
estimates (Table 6) to date give a value that is smaller
by two or three orders of magnitude. Therefore,
various processes for strong and fast dispersal of the
original cloud are suggested.
\begin{table}[ht]

\vspace{6mm}
\centering
{{\bf Table 6.} Statistical mass estimates for the Kuiper belt
}
\vspace{5mm}\begin{tabular}{l|c|c|c}
\hline \hline
\noalign{\smallskip}
\hline
\noalign{\smallskip}
Year & Author & Mass & Note \\
\noalign{\smallskip}
\hline
\noalign{\smallskip}
1997 & Weissman and Levison & 0.1 $\div$ 0.3 $m_{\oplus}$ & Between 30 au and 50 au \\
1998 & Jewitt et al. & $\sim 0.1$ $m_{\oplus}$ &  \\
1999 & Chiang and Brown & $\sim 0.2$ $m_{\oplus}$ & Between 30 au and 50 au \\
1999 & Kenyon and Luu & $\sim 0.1$ $m_{\oplus}$ & Between 30 au and 50 au \\
2001 & Gladman et al. & 0.04 $\div$ 0.1 $m_{\oplus}$ & Between 30 au and 50 au \\
2002 & Luu and Jewitt & 0.01 $\div$ 0.1 $m_{\oplus}$ & Between 35 au and 150 au \\
2002 & Kenyon & 0.1 -- 0.2 $m_{\oplus}$ & Total mass beyond the orbit Neptune \\  
2004 & Bernstein et al. & 0.010 $m_{\oplus}$ & Classic Kuiper belt \\
2009 & Booth et al. & 0.03 $m_{\oplus}$ & Class. + scattered Kuiper belt \\
 &  & 0.01 $m_{\oplus}$ & Classic objects \\
     &                    & 0.02 $m_{\oplus}$ & Scattered Kuiper objects \\
2010 & Vitense et al. & 0.05 $m_{\oplus}$ & Classic + resonant objects \\
     &                    & 0.07 $m_{\oplus}$ & Scattered objects \\
\hline
\end{tabular}
\end{table}
\subsection*{\it Dynamical Mass Estimates for the Kuiper Belt}
The Kuiper belt occupies a large volume of space
and includes not only a great number of large objects,
but also hundreds of thousands of smaller bodies. At
the current accuracy of planetary ephemerides the
attraction from the belt leads to a noticeable gravitational
influence on the motion of bodies in the Solar
system that should be properly taken into account.

The first dynamical mass estimates for TNOs in
the Kuiper belt were made in 2010 (Pitjeva 2010a,
2010b) based on the EPM2008 ephemerides. In this
case, the gravitational influence of a set of small or as yet undetected belt bodies was modeled by a one-dimensional material ring with a radius of 43 au in the plane of the ecliptic. The next mass estimates for
the Kuiper belt were made in a similar way using the
EPM2011 (Pitjeva 2013) and EPM2013 (Pitjeva and
Pitjev 2014) ephemerides. The results are presented
in Table 8. The data for the motion of Saturn obtained from Cassini radio measurements played an important role in the dynamical estimates found.
\begin{table}[ht]

\vspace{6mm}
\centering
{{\bf Table 8.} Dynamical mass estimates for the Kuiper belt}

\vspace{5mm}\begin{tabular}{l|c|c|c}
\hline \hline
Ephemerides & Mass of  & Total mass of belt & References\\          & 1D/2D ring &  ïîÿñà   &  \\
\hline
EPM2008 & $1.66\cdot10^{-2}\ m_{\oplus}$ & $2.58\cdot10^{-2}\ m_{\oplus}$ & Pitjeva(2010a, 2010b) \\ 
EPM2011 & $(1.67\pm 0.83)\cdot10^{-2}\ m_{\oplus}$ & $ 2.63\cdot10^{-2}\ m_{\oplus}$ & Pitjeva(2013)\\
EPM2013 & $(1.08 \pm 0.59)\cdot10^{-2}\ m_{\oplus}$ & $ 1.97\cdot10^{-2}\ m_{\oplus}$ & Pitjeva and Pitjev(2014)\\
EPM2016 & $(1.44 \pm 0.41)\cdot10^{-2}\ m_{\oplus}$ & $(2.28\pm 0.46)\cdot10^{-2}\ m_{\oplus}$ & Pitjeva et al.(2017)\\
EPM2017 & $(1.11 \pm 0.25)\cdot10^{-2}\ m_{\oplus}$ & $(1.97\pm 0.30)\cdot10^{-2}\ m_{\oplus}$ & This paper\\
\hline
\end{tabular}           	
\end{table}

In EPM2016 (Pitjeva et al. 2017) we made a transition
from modeling the total gravitational attraction
from small belt objects using a one-dimensional
ring (EPM2008--2015) to its modeling by a two-dimensional homogeneous ring. The densest part
of the Kuiper belt is the ring zone between the two
main 3 : 2 and 2 : 1 resonances with Neptune with the corresponding mean distances from the Sun $\sim$ 39.4 and $\sim$ 47.8 au. It contains the bulk of the Kuiper belt population and includes classical objects and the
most numerous part of resonant belt objects. Since
the number of objects outside this region drops significantly,
the distances of 39.4 and 47.8 au were taken
as, respectively, the inner and outer boundaries when modeling by the two-dimensional ring and when estimating
the mass of the Kuiper belt. For Saturn there are quite accurate observational data that were
obtained with the Cassini spacecraft and that are
presently very important for refining the gravitational
influence of the belt. The densest part of the belt
occupies a wide ring region (the radial width exceeds
8 au), and modeling its gravitational influence by
a two-dimensional ring makes it possible to more
properly take into account the influence on the planets nearest to it whose orbits are comparatively close to
the belt boundaries. It can be seen from our test
calculations (Table 7) that the influence of the two-dimensional ring differs noticeably from the influence of the one-dimensional ring with the same mass, especially on Uranus and Neptune.

\begin{table}[ht]

\vspace{6mm}
\centering
{{\bf Table 7.} Shift of the perihelion of planets due to the influence
of a homogeneous and two-dimensional ring over
100 years (in arcsec), $m=2\cdot10^{-2} m_{\oplus}$}

\vspace{5mm}\begin{tabular}{l|c|c|c}
\hline \hline
Planet &1 D&1 D&2 D\\
\cline{2-4}
 & R=43 au & R=44 au & $R_1=39.4$ au, $R_2=47.8$ au\\
\hline
  Neptune & $0^{\prime\prime}.0437$ & $0^{\prime\prime}.0376$ & $0^{\prime\prime}.0432$\\
 Uranus   & $0^{\prime\prime}.0095$ & $0^{\prime\prime}.0086$ & $0^{\prime\prime}.0091$\\
 Saturn & $0^{\prime\prime}.0024$ & $0^{\prime\prime}.0022$ & $0^{\prime\prime}.0023$\\
  Jupiter & $0^{\prime\prime}.0009$ & $0^{\prime\prime}.0008$ & $0^{\prime\prime}.0009$\\
  Mars   & $0^{\prime\prime}.0003$ & $0^{\prime\prime}.0003$ & $0^{\prime\prime}.0003$\\
\hline
\end{tabular}
\end{table}			
When using the two-dimensional ring as a model
(EPM2016, Pitjeva et al. 2017), we obtained the
following estimates:

the mass of the modeled TNOring
$$ M_{TNOring} =
(1.44 \pm 0.41) \cdot 10^{-2} \ m_{\oplus} \ (3 \sigma), $$ 
and the total mass of the Kuiper belt 
$$M_{Koiper}  =  (2.28 \pm 0.46)\cdot 10^{-2} \ m_{\oplus} \ (3 \sigma).$$ 
The dynamical masses determinations for the Kuiper belt are given in Table 8.

In this paper we obtained a new mass estimate
for the Kuiper belt using a {\bf discrete rotating model} for the attraction from small or as yet undetected belt
bodies and small fragments (the Section ``A Discrete
Rotating Model for the Belts''). Their combined
gravitational influence was modeled with the model
parameters $R_1 = 39.4,R_2 = 48.7,R_m = 44$ au and
a total number of material points of 160: we took 40
points for each of the $R_1$ = 39.4 and $R_2$ = 48.7 lines; for the densest part of the belt with a mean radius $R_m$ = 44 au we took 80 points.

The mass of the modeled part of the belt was found
to be
\begin{equation}\label{f-7}
 M_{TNOring} = (1.11 \pm 0.25) \cdot 10^{-2} \ m_{\oplus}\ \ (3 \sigma).  
 \end{equation}

Thus, the final result for {\bf the total mass of the belt},
including all of the large (Eq. (1)) and small (Eq. (2)) bodies, turned out to be
\begin{equation}\label{f-7}
   M_{Kuiper}= (1.97 \pm 030) \cdot 10^{-2} \ m_{\oplus}\qquad (3 \sigma) . 
\end{equation}

This estimate is given in the last row of Table 8.

As yet there are no dynamical mass estimates for
the Kuiper belt made by other authors.

\section*{\bf CONCLUSIONS}
We obtained new mass estimates for the main
asteroid belt and the Kuiper belt based on the revised
EPM2017 planetary ephemerides on the new ERA--
8 software platform and the software part for the
motion of the Moon. We used $\sim$ 800 000 positional
observations of planets and spacecraft (1913--2015).
To estimate the mass of the part of the belts that
consists of numerous small or as yet undetected belt
objects, we modeled the gravitational attraction using a discrete rotating system of material points. For the
main asteroid belt the total mass was found to be
$$ M_{belt} = (12.038 \pm 0.087) \cdot 10^{-10} \ M_{sun}= 
       (4.008 \pm 0.029) \cdot 10^{-4} \ m_{\oplus} \ (3 \sigma). $$
The total mass for the Kuiper belt is
$$  M_{Kuiper}= (1.97 \pm 0.30) \cdot 10^{-2} \ m_{\oplus}\qquad (3 \sigma) .$$

The error in estimating both the modeled part of
the belts and the total mass of the asteroid and Kuiper
belts decreased considerably from 2006 to 2017. This
is due to an increase in the number of observations
and an improvement of their quality as well as a
refinement of the models for the belts. Applying
the discrete model led to a better representation of
the observations (Figs. 1--7) and a refinement of the orbital elements for the planets (Table 3).

\section*{\bf ACKNOWLEDGMENTS}
We thank D.A. Pavlov for the development of the
ERA software package that allowed us to take an important
step in developing the EPM ephemerides and
the EPM2017 version: including the additional relativistic
Lense--Thirring accelerations in the general
model to integrate the differential equations; improving
the determination of the Solar system barycenter;
modeling the Sun's accelerations; integrating
the isochronous derivatives. We are also grateful to
M.A. Bodunova for her calculations of the influence
of the model rings on the planets. This work was
supported by the ``Cosmos: Studies of Fundamental
Processes and Their Interrelationships'' Program
no. 28 of the Presidium of the Russian Academy of
Sciences.

\end{document}